\documentclass{PoS}

\PoS{PoS(LAT2005)205}

\title{$B_s$ meson excited states from the lattice }

\ShortTitle{$B_s$ meson excited states}

\author{UKQCD Collaboration}

\author{A.~M.~Green$^a$, J. Ignatius$^b$, M.~Jahma$^c$,
\speaker{J.~Koponen$^a$}, C.~McNeile$^d$ and C.~Michael$^d$\\
\llap{$^a$}University of Helsinki and Helsinki Institute of Physics, Helsinki, Finland\\
\llap{$^b$}CSC - Scientific Computing Ltd., Espoo, Finland\\
\llap{$^c$}Helsinki University of Technology, Espoo, Finland\\
\llap{$^d$}University of Liverpool, Liverpool, UK\\
        E-mail: \email{jonna.koponen@helsinki.fi}}

\abstract{
This is a follow-up to our earlier work \cite{GKMP1, GKMP2, GK2CMT} for the
energies and the charge (vector) and matter (scalar) distributions for S-wave
states in a heavy-light meson, where the heavy quark is static and the light
quark has a mass about that of the strange quark. We now study the radial
distributions of higher angular momentum states, namely P- and D-wave states.
In nature the closest equivalent of this heavy-light system is the $B_s$ meson.

The calculation is  carried out with dynamical fermions on a $16^3\times 32$
lattice with a lattice spacing of about 0.10~fm generated with the non-perturbatively
improved clover action.
It is shown that several features of the energies and
radial distributions are in qualitative agreement with what one expects from a
simple one-body Dirac equation interpretation.
}

\FullConference{XXIIIrd International Symposium on Lattice Field Theory\\
		 25-30 July 2005\\
		 Trinity College, Dublin, Ireland}

\begin{document}

\section{Energies}

\begin{figure}[b!]
\centering
  \includegraphics*[width=0.55\textwidth]{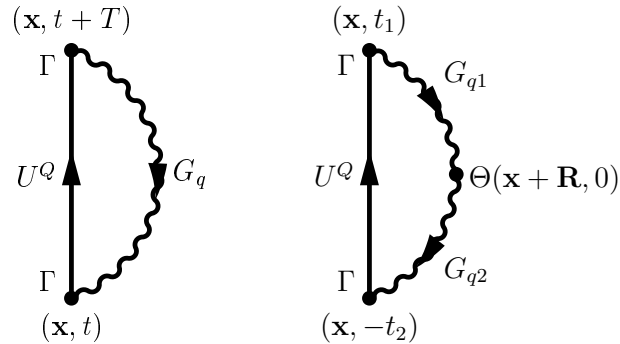}
\caption{The two-point correlation function $C_2$ (on the left) and
the three-point correlation function $C_3$ (on the right). The probe
$\Theta$ is at a distance $R$ from the heavy quark.}
\label{fig:C2C3}
\end{figure}

The basic quantity for evaluating the energies of heavy-light mesons is
the 2-point correlation function -- see Fig.~\ref{fig:C2C3}.
It is defined as
\begin{equation}
C_2(T)=\langle P_t\Gamma G_q(\mathbf{x},t+T,t)P_{t+T}
\Gamma^{\dag}U^Q(\mathbf{x},t,t+T)\rangle,
\end{equation}
where $U^Q(\mathbf{x},t,t+T)$ is the heavy (infinite mass)-quark propagator
and $G_q(\mathbf{x},t+T,t)$ the light anti-quark propagator. $P_t$
is a linear combination of products of gauge links at time $t$
along paths $P$ and $\Gamma$ defines the spin structure of the operator.
The $\langle ...\rangle$ means the average over the whole lattice.
The energies are then extracted by fitting the $C_2$
with a sum of exponentials,

%\begin{figure}
%\centering
%  \includegraphics*[width=0.20\textwidth]{diagramfig3.ps}
%\caption{The two-point correlation function $C_2$.}
%\label{fig:C2}
%\end{figure}

\begin{equation}
C_2(T)\approx\sum_{i=1}^{N_{\textrm{max}}}c_{i}\mathrm{e}^{-m_i T}c_i,
\end{equation}
where $N_{\textrm{max}}=2$ -- $4$, $T\leq 14$.

\begin{table}[b!]
\begin{center}
\begin{tabular}{|c|c|c|c|c|}
\hline
\hline
   &$\kappa$&$a$/fm&$m_q/m_s$&$r_0m_\pi$\\
\hline
DF3&0.1350  &0.110 &1.1      &1.93(3)   \\
DF4&0.1355  &0.104 &0.6      &1.48(3)   \\
DF5&0.1358  &0.099 &0.3      &1.06(3)   \\
\hline
\hline
\end{tabular}
\end{center}
\caption{Lattice parameters.
%The hadronic scale parameter is $r_0=0.525(25)$~fm.
 We use $r_0=0.525(25)$~fm to convert lattice results to physical units.
}
\label{LatParams}
\end{table}

\begin{figure}
\centering
 \includegraphics*[angle=-90,width=0.60\textwidth]{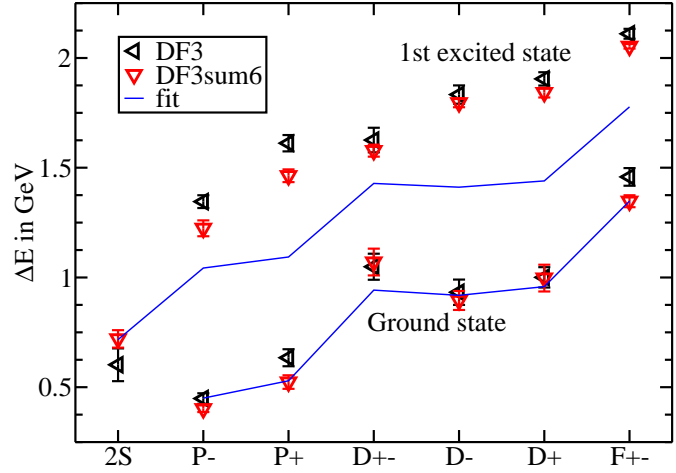}
\caption{The energies of different angular momentum states with respect
to the S-wave ground state. The error bars show the statistical error and
the estimated systematic error from the extraction procedure. No continuum
limit is taken.
% The energy spectrum for different angular momentum states.
% Here L$+$($-$) means that the light quark spin couples to orbital angular
% momentum L giving the total $j=L\pm 1/2$. 2S is the first radially
% excited $L=0$ state. Energies are given with respect to the S-wave
% ground state.
 See Section~\protect\ref{dirac} for more details about the Dirac model fit.
}
\label{fig:energies}
\end{figure}

 Calculations were made using three different lattices with $\beta=5.2$,
 $C_{\textrm{SW}}=2.0171$ non-perturbatively improved clover fermions.
 The parameters are given in Table~\ref{LatParams} and the extracted energies
 are summarized in Fig.~\ref{fig:energies}.
 The notation L$+$($-$) means that the light quark spin couples to orbital angular
 momentum L giving the total $j=\textrm{L}\pm 1/2$. 2S is the first radially
 excited L$=0$ state. Energies are given with respect to the S-wave
 ground state.
 The plot also shows a comparison
 between the static (infinitely heavy) and smeared (``sum6'') heavy quark.
 The ``sum6'' is APE type smearing,  where the simple gauge links in the time
 direction are replaced by the sum of six gauge link staples. The maximum distance
 from the original, simple link is one lattice spacing. The effect of the smearing
 on energy differences seems to be small.

\section{Radial distributions}

For evaluating the radial distributions of the light quark
a 3-point correlation function is needed -- see
Fig.~\ref{fig:C2C3}. It is defined as
\begin{equation}
C_3(R,T)=\langle \Gamma^{\dag}U^Q\Gamma G_{q1} \Theta
G_{q2}\rangle.
\end{equation}
We now have two light quark propagators, $G_{q1}$ and $G_{q2}$,
and a probe $\Theta(R)$ at distance $R$ from the
static quark. We have used two probes:
$\gamma_4$ for the vector (charge) and $1$
for the scalar (matter) distribution. The radial distributions, $x^{ij}(R)$,
are then extracted by fitting the $C_3$ with
\begin{equation}
C_3(R,T)\approx\sum_{i,j=1}^{N_{\textrm{max}}}c_{i}\mathrm{e}^{-m_i t_1}%
x^{ij}(R)\mathrm{e}^{-m_j t_2}c_{j},
\end{equation}
where the $m_i$, $c_i$ are those extracted from $C_2$.
The most recent calculations are the P$+$ and D$+-$ distributions
(see Figs.~\ref{figP}--\ref{figD}).
Earlier  S-wave distribution calculations have been published in
Refs.~\cite{GKMP1, GKMP2}.

%\begin{figure}
%\centering
%  \includegraphics*[width=0.20\textwidth]{diagramfig4.ps}
%\caption{The three-point correlation function $C_3$.}
%\label{fig:C3}
%\end{figure}

\begin{figure}
 \centering
 \includegraphics*[angle=-90,width=0.57\textwidth]{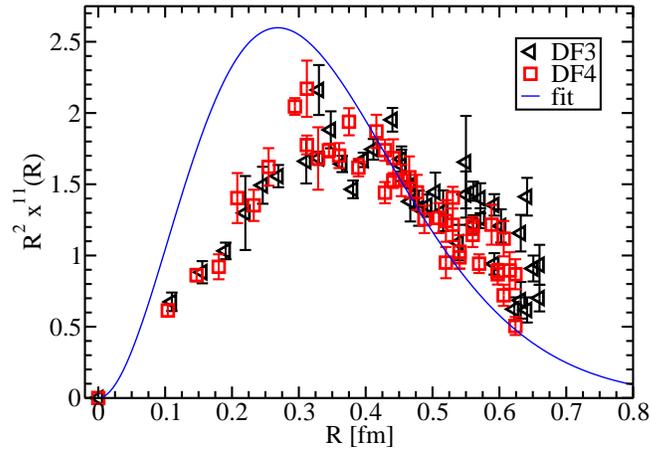}
\caption{
 S-wave ground state ($x^{11}$) charge distribution.}
\label{figS11}
\end{figure}

\begin{figure}
 \centering
 \includegraphics*[angle=-90,width=0.57\textwidth]{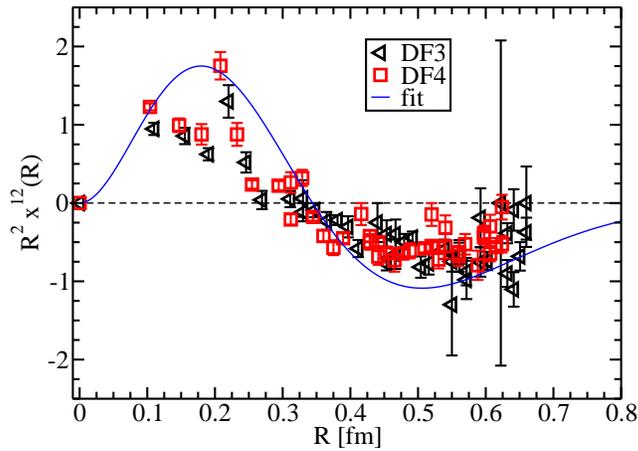}
\caption{
 S-wave charge distribution: the off-diagonal correlation between the ground state
 and the first radially excited state. The Dirac model seems to get
 the node right -- while no attempt has been made to fit the radial distributions.}
\label{figS12}
\end{figure}

%\begin{figure}
% \centering
% \includegraphics*[angle=-90,width=0.57\textwidth]{Swave_x13_R2_v2.ps.save}
%\caption{
% S-wave charge distribution: the off-diagonal correlation between the ground state
% and the second radially excited state. Two nodes are seen in the
% Dirac model fit, as expected, albeit shifted towards larger R.}
%\label{figS13}
%\end{figure}

\begin{figure}
 \begin{center}
\includegraphics*[angle=-90,width=0.57\textwidth]{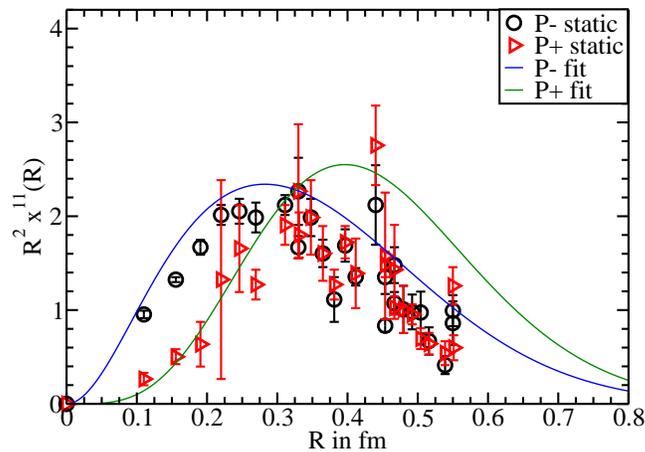}
\caption{P-wave charge distributions.}
\label{figP}
 \end{center}
\end{figure}

\begin{figure}
 \begin{center}
\includegraphics*[angle=-90,width=0.57\textwidth]{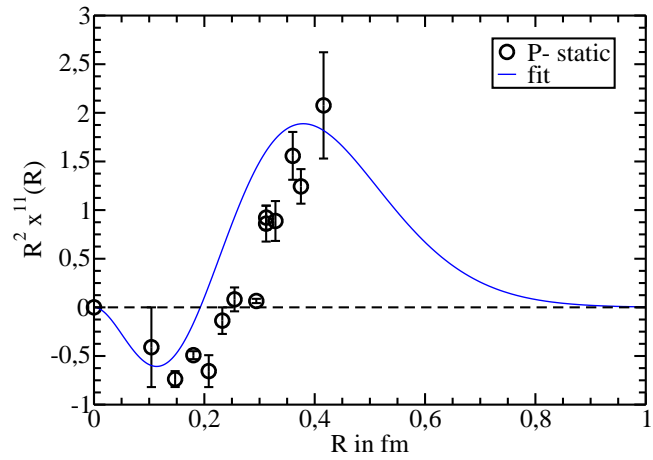}
\caption{P-wave matter distribution. Note that the Dirac model is able to get
the sign change right.}
\label{figPmatt}
 \end{center}
\end{figure}

\begin{figure}
 \centering
 \includegraphics*[angle=-90,width=0.57\textwidth]{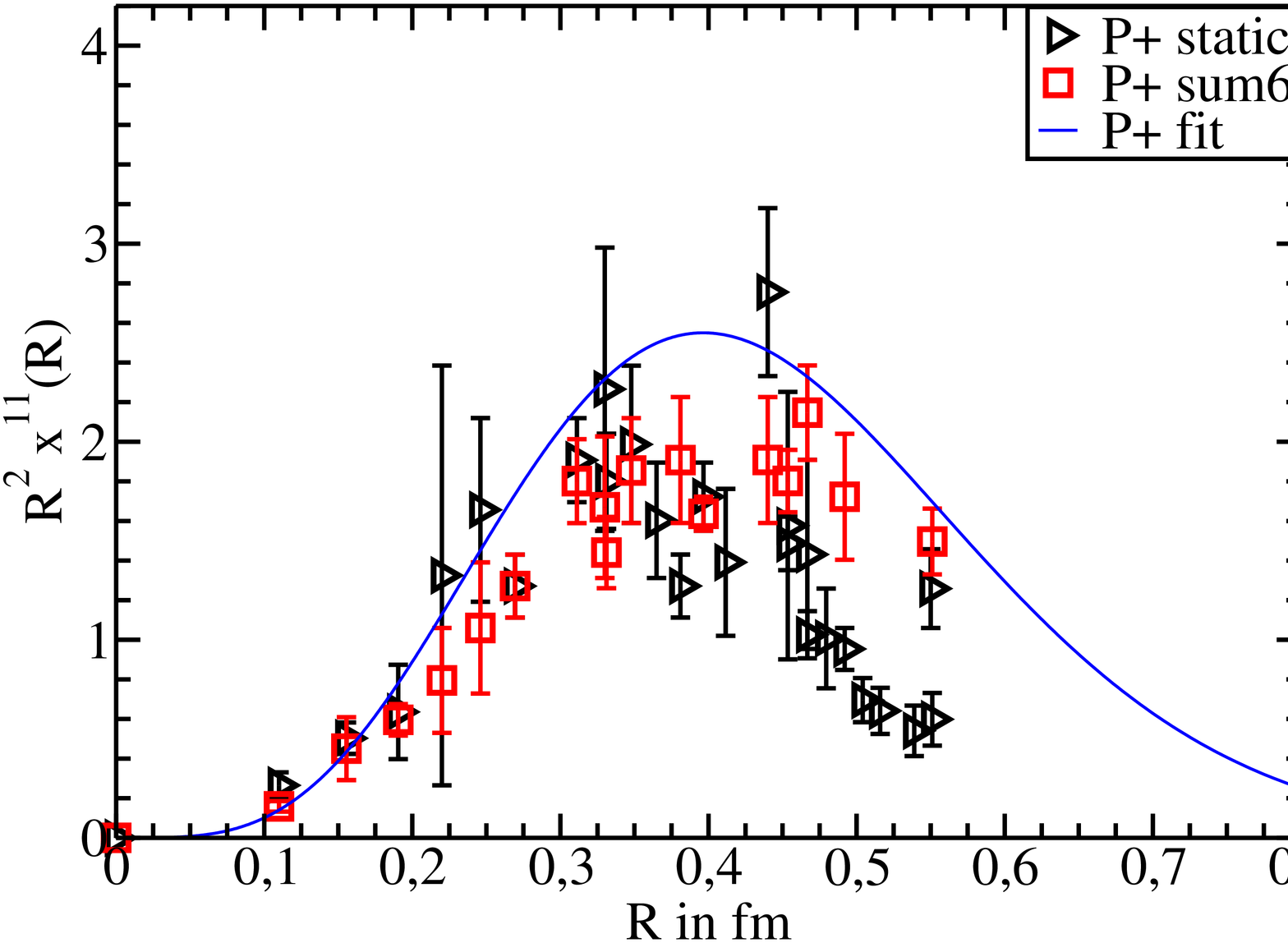}
\caption{Comparison of static (infinitely heavy) and smeared (``sum6'')
 heavy quark charge distributions for the P$+$ state.}
\label{figPsum6}
\end{figure}

\begin{figure}
 \centering
 \includegraphics*[angle=-90,width=0.57\textwidth]{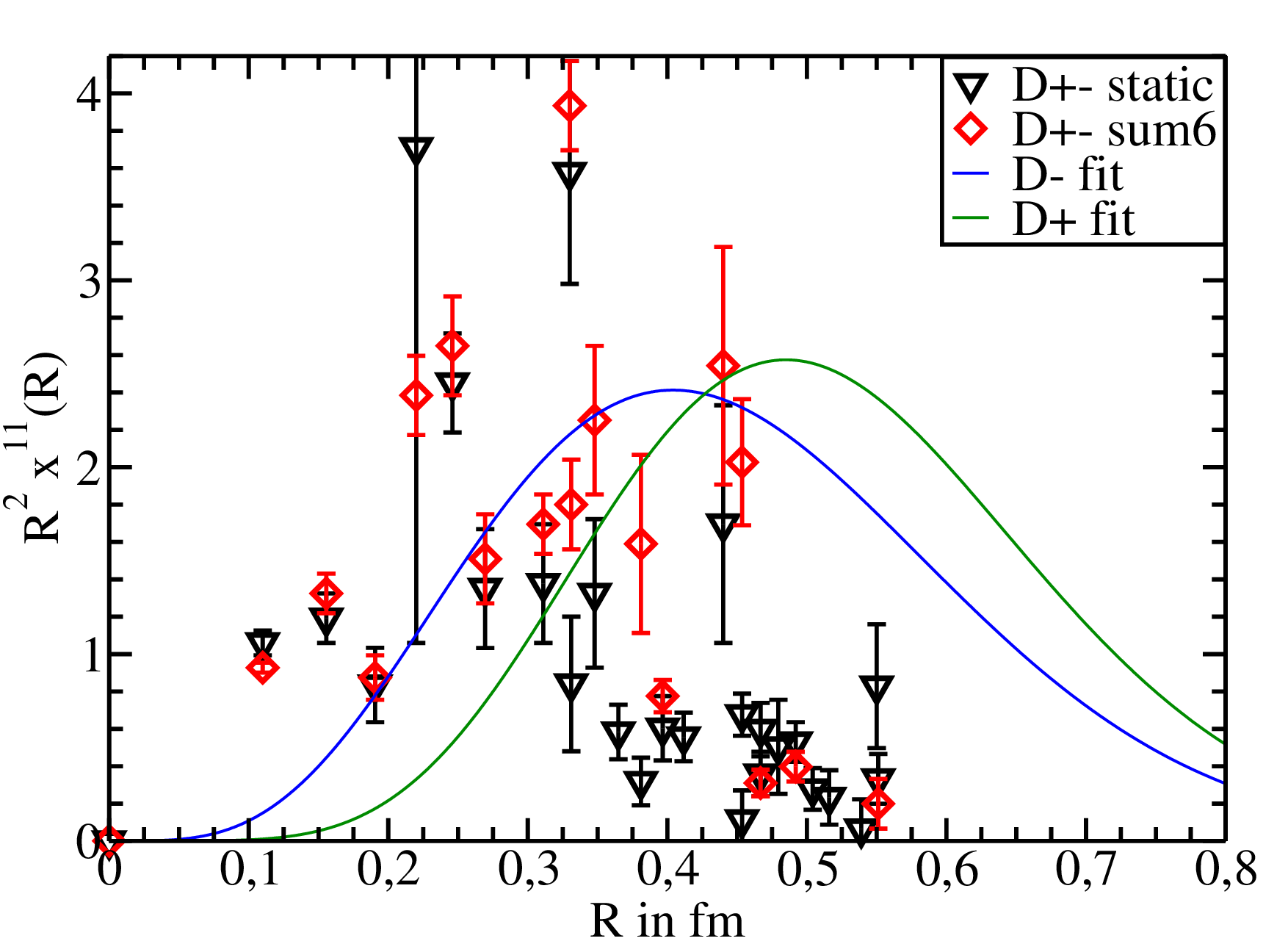}
\caption{Comparison of static (infinitely heavy) and smeared (``sum6'')
 heavy quark charge distributions for the D$+-$ state.% See also caption
% of Fig.~\protect\ref{figPsum6}.
}
\label{figD}
\end{figure}

\begin{figure}
 \centering
 \includegraphics*[angle=-90,width=0.547\textwidth]{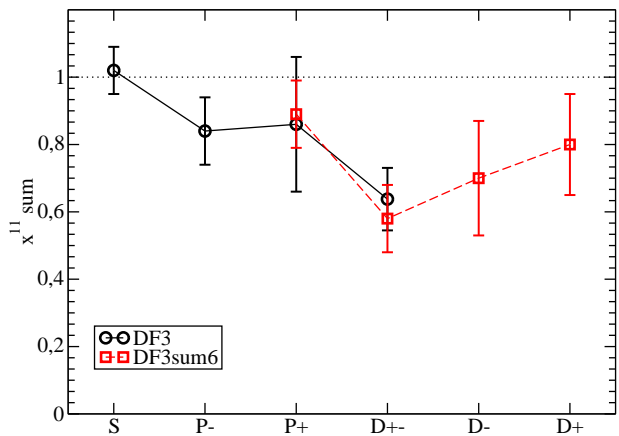}
\caption{Charge sum rule. This should be one (the charge of the $q$)
in the normalisation that is used here.}
\label{figSumrule}
\end{figure}

\section{A model based on the Dirac equation}
\label{dirac}

A simple model based on the Dirac equation is used to try
to describe the lattice data. Since the mass of the
heavy quark is infinite we have essentially a one-body
problem. The potential in the Dirac equation has a
linearly rising scalar part, $b_{\textrm{sc}} R$, as well as a
vector part $b_{\textrm{vec}} R$. The OGE potential, incorporating the
running coupling constant $\alpha_s(k^2)$, is obtained by the replacement
\begin{equation}
\frac{\alpha}{R}\;\longrightarrow\;
a_{\textrm{OGE}}\frac{2}{\pi}\int_0^{\infty}\!\! \mathrm{d}k\, j_0(kR)
\alpha_s(k^2),\quad \alpha_s(k^2)=\frac{12\pi}{27}
\ln^{-1}\frac{k^2+4 m^2_g}{\Lambda^2_{\textrm{QCD}}}.
\end{equation}
Here $\Lambda_{\textrm{QCD}}=260$~MeV, the dynamical gluon mass $m_g=290$~MeV
(see Ref.~\cite{Timo} for details) and $a_{\textrm{OGE}}$ is an overall adjustable parameter.
The potential has also a scalar term $m\omega \textrm{L}(\textrm{L}+1)$ depending on the orbital angular
momentum.

The solid lines in Figs. \ref{figS11}--\ref{figD} are radial
distributions from the Dirac model fit. The fit parameters are
$m = 0.088$~GeV, $a_{\textrm{OGE}} = 0.81$, $b_{\textrm{sc}} = 1.14$~GeV/fm,
$b_{\textrm{vec}} = 1.12$~GeV/fm and  $\omega = 0.028$. At present
we only fit energy differences --- the ground state energies and the 2S in
Fig.~\ref{fig:energies} --- and simply live with the resulting wavefunctions.
An attempt could be made to fit the radial distributions as well.

\section{Conclusions and acknowledgments}

%The key points are:
\begin{itemize}
\item
The spin-orbit splitting is small and supports the symmetry
$b_{\textrm{vec}} = b_{\textrm{sc}}$ as proposed in Ref.~\cite{Page}.
\item
There is now an abundance of lattice data to comprehend.
The energies and radial distributions of S, P and D$+-$ states can be
qualitatively understood by using a Dirac equation model.
\end{itemize}

J.~K. and A.~M.~G. wish to thank the UKQCD Collaboration for providing the
lattice configurations and the CSC - Scientific Computing Ltd.
for providing the computer resources. J.~K. and A.~M.~G. acknowledge
support by the Academy of Finland (contract 54038), the Magnus
Ehrnrooth foundation and the EU grant HPRN-CT-2002-00311 Euridice.
%J.K. thanks the Magnus Ehrnrooth foundation for financial support.

\end{document}